\documentclass[12pt]{article}
\usepackage{epsf}
\normalsize
\def\N{{\rm N}}
\def\bd{{\bf d}}

\def\bp{{\bf p}}
\def\bq{{\bf q}}
\def\bk{{\bf k}}

\def\cS{{\cal S}}
\def\tp{\tilde p}
\def\tP{\tilde P}
\def\tS{\tilde S}
\def\tq{\tilde q}
\def\tbp{\tilde\bp}
\def\tbq{\tilde\bq}

\def\Re{{\rm Re}}
\def\Im{{\rm Im}}

\def\lra{\leftrightarrow}
\def\alf{\alpha}
\def\eps{\epsilon}
\def\gam{\gamma}
\def\lam{\lambda}

\def\sg{\sigma}

\def\sgn{{\rm sgn}}
\def\det{{\rm det}}
\begin{document}
\title{Fermion loops, loop cancellation and density correlations
       in two dimensional Fermi systems}
\author{Arne Neumayr and Walter Metzner $^*$ \\
{\em Sektion Physik, Universit\"at M\"unchen} \\
{\em Theresienstra{\ss}e 37, D-80333 M\"unchen, Germany}}
\date{\small\today}
\maketitle
\renewcommand{\abstractname}{\normalsize{Abstract}}
\begin{abstract}
We derive explicit results from an exact expression for fermion loops 
with an arbitrary number of density vertices in two dimensions at zero
temperature, which
has been obtained recently by Feldman et al.\ \cite{FKST}.
The 3-loop is an elementary function of the three external momenta and
frequencies, and the N-loop can be expressed as a linear combination of
3-loops with coefficients that are rational functions of momenta and
frequencies.
We show that the divergencies of single loops for low energy and small
momenta cancel each other when loops with permuted external 
variables are summed.
The symmetrized N-loop, i.e.\ the connected N-point density correlation 
function of the Fermi gas, does not diverge for low energies and small
momenta.
In the dynamical limit, where momenta scale to zero at fixed finite
energy variables, the symmetrized N-loop vanishes as the 
$(2\N\!-\!2)$-th power of the scale parameter. 
\par\medskip
\noindent
\mbox{PACS: 71.10.-w, 71.10.Ca, 71.10.Pm}
\end{abstract}

\vfill\eject

\noindent
{\large\bf 1. Introduction} \par
\medskip

The discovery of high temperature superconductivity and the peculiar
phenomena observed in two-dimensional electron gases have stimulated much 
interest in the low energy physics of interacting Fermi systems with
singular interactions and reduced dimensionality.
In particular, the possible breakdown of Fermi liquid theory in such
systems has been a challenging issue \cite{MCD}.
\par\smallskip 
In this context the properties of fermion loops (see Fig.\ 1) with
density or current vertices play a very important role. 
Such loops appear as subdiagrams of Feynman diagrams and also as
kernels in effective actions for collective degrees of freedom.
Their behavior for small external momenta and energies determines
many properties of the whole system.
A single loop with N vertices diverges in the small energy-momentum 
limit for $\N>2$. 
In the symmetrized loop, obtained by summing all permutations of the 
external energy-momentum variables $q_1,\dots,q_\N$, at least the
leading contributions from single loops in the small energy-momentum
limit are known to cancel each other \cite{KHS,MCD,Kop}.
This systematic cancellation is known as {\em loop cancellation}.
As a consequence, loops with $\N > 2$ are often irrelevant for the
low-energy behavior, and can therefore be neglected in effective
actions or in resummations of the perturbation series via Ward 
identities or bosonization \cite{MCD,Kop}.
This implies that collective density fluctuations and the associate 
response functions can be described by a renormalized random
phase approximation, as in a Fermi liquid, even if interactions
destroy Landau quasi-particles.
\par\smallskip
While the 2-loop, known as polarization insertion or bubble diagram,
has been computed long ago in one, two and three dimensions, no
explicit formulae are available for higher order loops.
Furthermore, the available proofs of loop cancellation show only that
some cancellation occurs, but they do not show by how many powers
the divergence is reduced.
To be able to carry out an accurate power-counting of the infrared
divergencies of systems with singular interactions or soft bosonic 
modes, a more detailed knowledge of the behavior of loops is required.
\par\smallskip
Recently, Feldman, Kn\"orrer, Sinclair and Trubowitz \cite{FKST} have
obtained an exact expression for the N-loop with density vertices in a 
two dimensional Fermi gas at zero temperature. 
In this article we derive explicit formulae for loops with density
vertices in two dimensions from their expression, which are useful for 
the evaluation of Feynman diagrams containing such loops as subdiagrams.
We show that loop cancellation
reduces the degree of divergence by $\N\!-\!2$ powers for $\N>2$, i.e.\
the symmetrized loops, which are equal to the N-point density correlation
functions of the Fermi gas, are generically {\em finite}\/ in the small 
energy-momentum limit.
In the dynamical limit, where the momenta scale to zero at fixed finite
energy variables, the symmetrized N-loop vanishes as the $(2\N\!-\!2)$-th 
power of the scale parameter.
\par\smallskip
This article is organized as follows. 
In Sec.\ 2 we define single and symmetrized loops and review briefly
known basic properties. 
In Sec.\ 3 we summarize the main results on loops in two dimensions
obtained by Feldman et al.\ \cite{FKST}, which are
the starting point for our analysis.
In Sec.\ 4 the 3-loop is evaluated explicitly, and its asymptotic behavior 
for small energy and momentum variables is analysed.
In Sec.\ 5 we derive a theorem clarifying the structure of a certain
symmetrized product, which allows us to control the asymptotic low
energy and small momentum behavior of symmetrized N-loops, to be 
discussed in Sec.\ 6. 
A conclusion follows in Sec.\ 7.
\begin{figure}
\epsfbox{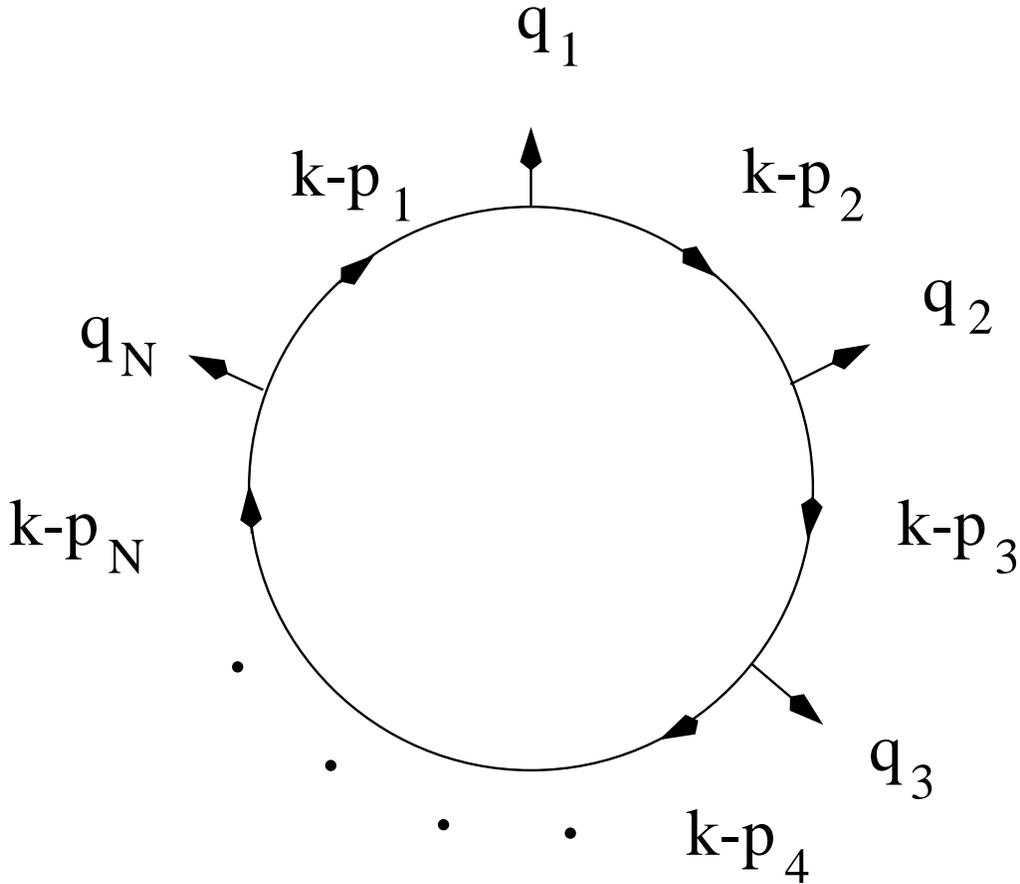}
\caption{The N-loop with its energy-momentum labels.}
\end{figure}
\par\vskip 0.7 cm

\noindent
{\large\bf 2. Definitions and simple properties of loops} 
\par\medskip

The amplitude of the N-loop with density vertices, represented by the 
Feynman diagram in Fig.\ 1, is given by
\begin{equation}\label{eq1}
 \Pi_\N(q_1,\dots,q_\N) = I_\N(p_1,\dots,p_\N) =  
 \int\! \frac{d^dk}{(2\pi)^d} \int\! \frac{dk_0}{2\pi} \, 
 \prod_{j=1}^\N  G_0(k\!-\!p_j)
\end{equation}
at temperature zero.
Here $k = (k_0,\bk)$, $q_j = (q_{j0},\bq_j)$ and $p_j = (p_{j0},\bp_j)$
are energy-momentum vectors.
The variables $q_j$ and $p_j$ are related by the linear transformation
\begin{equation}\label{eq2}
 q_j = p_{j+1} - p_j \>, \quad j = 1,\dots,\N
\end{equation}
where $p_{\N+1} \equiv p_1$. 
Energy and momentum conservation at all vertices
yields the restriction $q_1 + \dots + q_\N = 0$.
The variables $q_1,\dots,q_\N$ fix $p_1,\dots,p_\N$ only up to a constant
shift $p_j \mapsto p_j + p$. Setting $p_1 = 0$, one gets
\begin{eqnarray}\label{eq3}
 p_2 & = & q_1 \nonumber \\
 p_3 & = & q_1 + q_2 \nonumber \\
     & \vdots & \nonumber \\
 p_\N & = & q_1 + q_2 + \dots + q_{\N-1} 
\end{eqnarray}
We use the imaginary time representation, with a 
non-interacting propagator
\begin{equation}\label{eq4}
 G_0(k) = \frac{1}{ik_0 - (\eps_{\bk} - \mu)} 
\end{equation}
where $\eps_{\bk}$ is the dispersion relation and $\mu$ the chemical
potential of the system.
The $k_0$-integral in Eq.\ (\ref{eq1}) can be easily carried out using the
residue theorem; one obtains \cite{FKST}
\begin{equation}\label{eq5}
 I_\N(p_1,\dots,p_n) = \sum_{i=1}^N \int_{|\bk-\bp_i| < k_F}
 \frac{d^dk}{(2\pi)^d} 
 \left( \prod_{j=1 \atop j \neq i}^n f_{ij}(\bk) \right)^{-1}
\end{equation}
where 
$f_{ij}(\bk) = \eps_{\bk-\bp_i} - \eps_{\bk-\bp_j} + i(p_{i0} - p_{j0})$.
\par\smallskip
The 2-loop $\Pi_2(q,-q) \equiv \Pi(q)$ is known as polarization insertion
or particle-hole bubble, and has a direct physical meaning: $\Pi(q)$ is
the dynamical density-density correlation function of a non-interacting
Fermi system \cite{FW}.
While the 2-loop has been computed explicitly for continuum systems
in one, two and three dimensions already long ago \cite{MCD}, only few
explicit results exist for higher loops.
\par\smallskip
The behavior of loops for small $q_j$ is particularly important,
because interactions and bosonic propagators attached to the fermion
loops in general Feynman diagrams are often singular for small energies
and momenta \cite{MCD}.
\par\smallskip 
A simple formula for the {\em static}\/ small-q limit has been derived
long ago by Hertz and Klenin \cite{HK}, namely
\begin{equation}\label{eq6}
 \lim_{\bq_j \to 0} \lim_{q_{j0} \to 0} \Pi_\N(q_1,\dots,q_\N) =
 \frac{(-1)^{\N-1}}{(\N\!-\!1)!} 
 \left. \frac{d^{\N-2} D(\eps)}{d\eps^{\N-2}} \right|_{\eps=\mu}
\end{equation}
where 
$D(\eps) = \int \frac{d^dk}{(2\pi)^d} \, \delta(\eps\!-\!\eps_{\bk})$ 
is the density of states.
Note that the limit $q_j \to 0$ is not unique; the above result is
valid only in the so-called static limit where the energy variables
go to zero first.
\par\smallskip
In the following we will analyze the behavior of loops in the generic
small-q limit, $\lim_{\lam \to 0} \Pi_\N(\lam q_1,\dots,\lam q_\N)$ for
arbitrary $q_1,\dots,q_\N$, with a vanishing scaling factor $\lam$
applied to all energy and momentum variables.
According to a simple power-counting estimate of the integral in 
Eq.\ (\ref{eq1}) one would expect that an N-loop with $N>2$ diverges as 
$\lam^{2-N}$ for $\lam \to 0$,
since for $q_j = 0$ one integrates over an N-fold divergence on the
$(d-1)$-dimensional manifold defined by $k_0 = 0$, $\eps_{\bk} = \mu$, 
which has codimension two in a $(d+1)$-dimensional energy-momentum 
space.
This estimate yields however only an upper bound for the degree of
divergence, since the actual value of the integral may be smaller
due to cancellations of contributions with opposite signs, as is
obviously the case in the static limit (\ref{eq6}).
\par\smallskip
We will also analyze the so-called {\em dynamical}\/ limit, where a 
vanishing scaling factor $\lam$ is applied only to momentum variables, 
at fixed finite energy variables.
\par\smallskip
Divergencies of single loops in the small-q limit cancel each other
at least to some degree in the {\em symmetrized}\/ loop
\begin{equation}\label{eq7}
 \Pi^S_\N(q_1,\dots,q_\N) =
 \cS \, \Pi_\N(q_1,\dots,q_\N) =
 \frac{1}{\N!} \sum_P \Pi_\N(q_{P1},\dots,q_{P\N})
\end{equation}
where the symmetrization operator $\cS$ imposes summation over all 
permutations of $q_1,\dots,$ $q_\N$.
The behavior of single loops is physically not relevant, since all
physical properties can be expressed in terms of symmetrized loops.
This important point is however ignored in some approximation
schemes.
The symmetrized N-loop is proportional to the N-point density
correlation function of a non-interacting Fermi system with dispersion
relation $\eps_{\bk}$ and chemical potential $\mu$.
In the one-dimensional Luttinger model \cite{Voi}, which has a linear 
dispersion relation, loops with $N>2$ cancel completely, i.e. 
$\Pi_\N^S \equiv 0$, as first noticed by Dzyaloshinskii and Larkin 
\cite{DL}.
In other systems the loop cancellation is not complete and, in general,
$\Pi_\N^S$ does not vanish in the small-q limit. Only in the
dynamical limit Ward identities associated with particle number 
conservation imply that $\Pi_N^S$ goes to zero at least linearly in
each momentum variable \cite{CDM,Tru}. 
This is crucial in particular for the infrared renormalizability of the 
Coulomb gas \cite{Tru}, since the small-q singularity of
the interaction is not removed by screening in the dynamical limit.
For the general small-q limit with finite ratios of momenta and energies,
only the cancellation of leading singularities has already been 
established \cite{KHS,MCD,Kop}, but to our knowledge the behavior of the 
remainder has not yet been analyzed.
\par

\vskip 1cm

\noindent
{\large\bf 3. Two dimensions and the results by Feldman et al.} \par
\medskip 

We now focus on {\em two-dimensional}\/ Fermi systems with a quadratic 
dispersion relation $\eps_{\bk} = \bk^2/2m$, where $m$ is the mass of
the particles. The Fermi surface is then a circle with radius
$k_F = \sqrt{2m\mu}$. In the following we choose energy and momentum
units such that $k_F = 1$ and $m = 1$. It is easy to reinsert arbitrary
values for $k_F$ and $m$ by making a simple dimensional analysis of the
expressions.
\par\smallskip
The 2-loop in two dimensions, first computed by Stern \cite{Ste}, is
given by
\begin{equation}\label{eq8}
 \Pi(q) = \frac{1}{2\pi} \left\{ -1 +
 \frac{1}{|\bq|} \left[ 
  \sqrt{\left(-\frac{iq_0}{|\bq|}+\frac{|\bq|}{2}\right)^2 - 1} +
  \sqrt{\left(\frac{iq_0}{|\bq|}+\frac{|\bq|}{2}\right)^2 - 1} \,
  \right] \right\}  
\end{equation}
where the complex square-root is defined to have positiv real part.
In the small-q limit one obtains
\begin{equation}\label{eq9}
 \lim_{\lam \to 0} \Pi(\lam q) =
 \frac{1}{2\pi} \left( -1 + \frac{1}{\sqrt{1 + |\bq|^2/q_0^2}} \right)
\end{equation}
i.e.\ a finite result depending on the ratio $|\bq|/q_0$.
In the dynamical limit, $\bq \to 0$ at finite $q_0$, the 2-loop vanishes
as $|\bq|^2$.
\par\smallskip
Using methods from complex analysis,
Feldman et al.\ \cite{FKST} have recently reduced the problem of computing
arbitrary N-loops in two dimensions to elementary integrals.
We now report their results, which are the starting point for our
further explicit evaluations.
\par\smallskip
If the three vectors $\bp_i$, $\bp_j$ and $\bp_k$ are not collinear,
one can define 
the two-dimensio\-nal complex vector $\bd^{ijk}$ as the unique 
solution of the equations $f_{ij}(\bk) = f_{jk}(\bk) = 0$, where
\begin{equation}\label{eq10}
 f_{ij}(\bk) = (\bp_j - \bp_i)\cdot\bk + 
 \frac{1}{2}(\bp_i^2 - \bp_j^2) + i(p_{i0} - p_{j0})
\end{equation}
Since $f_{ij} + f_{jk} + f_{ki} \equiv 0$, the vector 
$\bd^{ijk}$ is also a solution of the equation $f_{ki}(\bk) = 0$.
Hence, $\bd^{ijk}$ does not depend on the order of the indices $i,j,k$.
The real part of $\bd^{ijk}$ is the center of the circle circumscribing
the triangle with vertices $\bp_i,\bp_j,\bp_k$.
\par\smallskip
For the 3-loop, Feldman et al.\ \cite{FKST} have obtained the expression
\begin{equation}\label{eq11}
 I_3(p_1,p_2,p_3) = 
 \frac{1}{2\pi i \, \det(\bp_2\!-\!\bp_1,\bp_3\!-\!\bp_1)}
 \sum_{i,j=1 \atop i \neq j}^3 s_{ij} \int_{\gam_{ij}} \frac{dz}{z}
\end{equation}
where $s_{12} = s_{23} = s_{31} = 1$, $s_{21} = s_{32} = s_{13} = -1$,
and $\det({\bf u},{\bf v}) \equiv u_x v_y - u_y v_x$ for arbitrary 
two-dimensional vectors ${\bf u}, {\bf v}$.
The contour-integrals are performed along the curves
$\gam_{ij} = \{ w_{ij}(s) | 0 \leq s \leq 1 \}$
where $w_{ij}(s)$ is the unique (generally complex) root of the quadratic
equation
\begin{equation}\label{eq12}
 (\bp_j\!-\!\bp_i)^2 \, z^2 + 
 2 \, \det(\bd\!-\!\bp_i,\bp_j\!-\!\bp_i) \, z +
 (\bd\!-\!\bp_i)^2 = s^2
\end{equation}
satisfying the condition
\begin{equation}\label{eq13}
 \Im \, \frac{-(p_{jx} - p_{ix}) w_{ij}(s) + d_y - p_{iy}}
          { (p_{jy} - p_{iy}) w_{ij}(s) + d_x - p_{ix}} \> > \> 0
\end{equation}
with $\bd = (d_x,d_y) = \bd^{123}$.
\par\smallskip
The general N-loop can also be written as a complex contour integral 
over rational functions \cite{FKST}. That expression implies that the 
N-loop can be expressed in terms of 3-loops as
\begin{equation}\label{eq14}
 I_\N(p_1,\dots,p_\N) = \sum_{1 \leq i < j < k \leq \N}
 \left[ \prod_{\nu = 1 \atop \nu \neq i,j,k}^\N 
 f_{i\nu}(\bd^{ijk}) \right]^{-1} I_3(p_i,p_j,p_k)
\end{equation}
This {\em reduction formula}\/ follows also directly from the identity
\begin{equation}\label{eq15}
 \prod_{i=1}^\N G_0(k\!-\!p_i) = \sum_{1 \leq i < j < k \leq \N}
 \left[ \prod_{\nu = 1 \atop \nu \neq i,j,k}^\N 
 f_{i\nu}(\bd^{ijk}) \right]^{-1} 
 G_0(k\!-\!p_i) G_0(k\!-\!p_j) G_0(k\!-\!p_k)
\end{equation}
valid for $G_0(k) = [ik_0 - (\bk^2/2 - \mu)]^{-1}$ in two dimensions
\cite{CW}.
Note that 
$f_{i\nu}(\bd^{ijk}) = f_{j\nu}(\bd^{ijk}) = f_{k\nu}(\bd^{ijk})$.
\par\smallskip
A remarkable explicit result has been obtained by Feldman et al.\ 
\cite{FKST} in the static case ($q_{i0} = 0$ for all $i$). 
In that case they show that $I_\N$ vanishes identically for $\N > 2$
when all disks with radius $k_F$ around the points $\bp_1,\dots,\bp_\N$
have at least one point in common!
This implies that $\Pi_\N(q_1,\dots,q_\N)$ vanishes identically for 
$q_{i0} = 0$ and small $\bq_i$, if $\N > 2$, which is compatible with
Eq.\ (\ref{eq6}) since $D(\eps)$ is constant in two dimensions.
\par\smallskip
In the following we will derive more explicit formulae for 3-loops,
and will analyze the behavior of $\Pi_\N$ and $\Pi^S_\N$ in various
important limits. 
To this end, it is useful to determine the point $\bd^{ijk}$ explicitly
as a function of $p_i,p_j,p_k$. Writing the inhomogeneous linear system
of equations for $\bd^{ijk}$ in matrix form and inverting the matrix,
one obtains
\begin{equation}\label{eq16}
 \bd^{ijk} = \frac{1}{\det(\bp_j\!-\!\bp_i,\bp_k\!-\!\bp_i)} \, 
  \left[ \frac{1}{2} (\bp_k^2 - \bp_i^2) + i(p_{k0} - p_{i0}) \right]
  (\bp_j - \bp_i)^{\perp} \> + \> j \leftrightarrow k 
\end{equation}
Here $\bp^{\perp} = (p_x,p_y)^{\perp} = (-p_y,p_x)$ denotes a rotation
by an angle $\pi/2$ and $j \leftrightarrow k$ means exchange of indices
$j$ and $k$.
Note that $|\det(\bp_j\!-\!\bp_i,\bp_k\!-\!\bp_i)|$ is twice the area 
of the triangle with vertices $\bp_i,\bp_j,\bp_k$.
If $\bp_i,\bp_j,\bp_k$ are collinear, this area vanishes and the point 
$\bd^{ijk}$ tends to infinity.
Using (\ref{eq16}), the coefficient $f_{i\nu}(\bd^{ijk})$ appearing in
the reduction formula (\ref{eq14}) can be expressed explicitly as
\begin{eqnarray}\label{eq17}
 f_{i\nu}(\bd^{ijk}) &= &
 \frac{1}{2} (\bp_i^2 - \bp_{\nu}^2) + i(p_{i0} - p_{\nu 0}) \nonumber \\
 &+ & \left\{ 
  \left[ \frac{1}{2} (\bp_k^2 - \bp_i^2) + i(p_{k0} - p_{i0}) \right]
  \frac{\det(\bp_j\!-\!\bp_i,\bp_{\nu}\!-\!\bp_i)}
       {\det(\bp_j\!-\!\bp_i,\bp_k\!-\!\bp_i)} \> + \>
  j \leftrightarrow k
 \right\}
\end{eqnarray}
We conclude this section by defining a complex function of $p_i,p_j,p_k$ 
that will be useful in the following evaluations:
\begin{equation}\label{eq18}
 \bar z_{ijk} = \bar x_{ijk} + i \bar y_{ijk} =
 \frac{\det(\bp_j\!-\!\bd^{ijk},\bp_i\!-\!\bd^{ijk})}{|\bp_j\!-\!\bp_i|} =
 \det\left(\bd^{ijk}\!-\!\bp_i,\frac{\bp_j\!-\!\bp_i}{|\bp_j\!-\!\bp_i|}
     \right)
\end{equation}
Inserting $\bd^{ijk}$, one obtains
\begin{eqnarray}\label{eq19}
 \bar x_{ijk} &=& 
  \frac{|\bp_j\!-\!\bp_i|}{2 \, \det(\bp_j\!-\!\bp_i,\bp_k\!-\!\bp_i)}
  \, (\bp_j\!-\!\bp_k)\cdot(\bp_k\!-\!\bp_i)
 \nonumber \\
 \bar y_{ijk} &=&
 \frac{1}{\det(\bp_j\!-\!\bp_i,\bp_k\!-\!\bp_i)} \,
 \frac{\bp_j\!-\!\bp_i}{|\bp_j\!-\!\bp_i|} \cdot 
 \left[ (\bp_k\!-\!\bp_i) (p_{j0}\!-\!p_{k0}) -
        (\bp_j\!-\!\bp_k) (p_{k0}\!-\!p_{i0})
 \right]
\end{eqnarray}
The modulus of $\bar x_{ijk}$ is twice the distance between the point
$\Re\,\bd^{ijk}$ and the straight line connecting $\bp_i$ and $\bp_j$.
Note the useful identity
\begin{equation}\label{eq20}
 (\bd^{ijk}\!-\!\bp_i)^2 - \bar z_{ijk}^2 =
 \left[ \frac{1}{2} |\bp_j\!-\!\bp_i| + 
    i \,\frac{p_{j0}\!-\!p_{i0}}{|\bp_j\!-\bp_i|} \right]^2
\end{equation}

\vskip 1cm

\noindent
{\large\bf 4. The 3-loop} \par
\medskip
In this section we derive an explicit formula for the 3-loop and
analyze its behavior in various important limits.
\par\smallskip
We first derive an explicit expression for the functions $w_{ij}(s)$,
$i,j \in \{1,2,3\}$, $i \neq j$.
Equation (\ref{eq12}) can also be written as
\begin{equation}\label{eq21}
 \left[ |\bp_j\!-\!\bp_i| z + \bar z_{ij} \right]^2 +
 (\bd\!-\!\bp_i)^2 - \bar z_{ij}^2 = s^2
\end{equation}
where 
$\bar z_{ij} = \bar x_{ij} + \bar y_{ij} \equiv \bar z_{ijk}$ 
with $k \in \{1,2,3\}$, $k \neq i,j$.
Making the ansatz
\begin{equation}\label{eq22}
 w_{ij}(s) = 
 \frac{z_{ij}(s) - \bar z_{ij}}{|\bp_j\!-\bp_i|}
\end{equation}
with a complex function $z_{ij}(s) = x_{ij}(s) + iy_{ij}(s)$,
and using relation (\ref{eq20}), one obtains the equation
\begin{equation}\label{eq23}
 z_{ij}^2(s) + \left[ \frac{1}{2} |\bp_j\!-\!\bp_i| + 
    i \,\frac{p_{j0}\!-\!p_{i0}}{|\bp_j\!-\bp_i|} \right]^2 = s^2
\end{equation}
Splitting real and imaginary parts, one obtains two real equations for 
$ x_{ij}(s)$ and $ y_{ij}(s)$,
\begin{eqnarray}\label{eq24}
 x_{ij}^2(s) - y_{ij}^2(s) &=&
 s^2 - \frac{1}{4} |\bp_j\!-\!\bp_i|^2 + 
       \frac{(p_{j0}\!-\!p_{i0})^2}{|\bp_j\!-\!\bp_i|^2} 
 \> \equiv \> a_{ij}(s) \nonumber \\
  x_{ij}(s) \, y_{ij}(s) &=&
 - \frac{1}{2} (p_{j0}\!-\!p_{i0})
\end{eqnarray}
and the condition (\ref{eq13}) implies
\begin{equation}\label{eq25}
 (p_{j0}\!-\!p_{i0}) \, x_{ij}(s) -
 \frac{1}{2} |\bp_j\!-\!\bp_i|^2 \, y_{ij}(s) \> > \> 0
\end{equation}
The two equations (\ref{eq24}) define intersecting hyperbolas in the
$(x,y)$-plane. 
Note that the second equation does not depend on the variable $s$. 
For any $s$ there is exactly one intersection point satisfying the
condition (\ref{eq25}), given by
\begin{eqnarray}\label{eq26}
 x_{ij}(s) & = &
 \sgn(p_{j0}\!-\!p_{i0})  
 \frac{1}{\sqrt{2}} \, 
 \sqrt{\sqrt{[a_{ij}(s)]^2 + (p_{j0}\!-\!p_{i0})^2} + a_{ij}(s)}
 \nonumber \\
  y_{ij}(s) & = & 
 - \, \frac{1}{\sqrt{2}} \, 
 \sqrt{\sqrt{[a_{ij}(s)]^2 + (p_{j0}\!-\!p_{i0})^2} - a_{ij}(s)}
\end{eqnarray}
All the square-roots in (\ref{eq26}) have real positive arguments.
\par\smallskip
It is now obvious that the curve 
$\gam_{ij} = \{ w_{ij}(s) | 0 \leq s \leq 1 \}$ is connected and
sweeps out an angle smaller than $\pi$ with respect to the origin in 
the complex plane.
Hence
\begin{equation}\label{eq27}
 \int_{\gam_{ij}} \frac{dz}{z} = 
 \ln\left[\frac{w_{ij}(1)}{w_{ij}(0)}\right]
\end{equation}
where $\ln$ is the main-branch of the complex logarithm, which is real
and continuous on the positive real axis and has a branch cut on the
negative real axis. Hence,
\begin{equation}\label{eq28}
 I_3(p_1,p_2,p_3) = 
 \frac{1}{2\pi i \, \det(\bp_2\!-\!\bp_1,\bp_3\!-\!\bp_1)}
 \sum_{i,j=1 \atop i \neq j}^3 s_{ij}
 \ln\left( 
  \frac{z_{ij}(1) - \bar z_{ij}}
       {z_{ij}(0) - \bar z_{ij}}
      \right)
\end{equation}
We have thus obtained an explicit formula for the 3-loop as a function
of $p_1,p_2,p_3$, which is useful in particular for the evaluation of
Feynman diagrams with 3-loops as subdiagrams. 
Note that the result $I_3 = 0$ in the static limit (for momenta such
that unit disks around them have at least one point in common)
is not obvious from Eq.\ (\ref{eq28}), and is obtained only as a 
consequence of cancellations in the sum over $i,j$.
\par\smallskip
We now analyze the behavior of single 3-loops and the symmetrized 3-loop
in the small-q limit defined in Sec.\ 2. 
With the choice $p_1 = 0$, the substitution $q_i \mapsto \lam q_i$
implies $p_i \mapsto \lam p_i$. 
The function $\bar y_{ij}$ is invariant under this substitution, while
$\bar x_{ij}$ becomes a homogeneous function of $\lam$ of order one,
i.e. $\bar x_{ij} \mapsto \lam \bar x_{ij}$.
For $\lam \to 0$, the functions $x_{ij}(s)$ and $y_{ij}(s)$ 
reduce to
\begin{eqnarray}\label{eq29}
  x_{ij}(s) & \to &
 \sgn(p_{j0}\!-\!p_{i0}) \sqrt{s^2 + 
 (p_{j0}\!-\!p_{i0})^2/|\bp_j\!-\!\bp_i|^2}  \nonumber \\
  y_{ij}(s) & \to &
  - \,\frac{\lam}{2} \, \frac{|\bp_j\!-\!\bp_i|} 
  {\sqrt{1 + s^2 |\bp_j\!-\!\bp_i|^2/(p_{j0}\!-\!p_{i0})^2}}
\end{eqnarray}
Hence $x_{ij}(s)$ is of order one for $\lam \to 0$ while
$y_{ij}(s)$ is of order $\lam$.
To avoid confusion, we emphasize that we consider the small-q limit
for {\em finite}\/ fixed ratios between the various momenta and 
frequencies. The following small-q expansion is valid only if 
energy/momentum ratios $\frac{p_{j0}-p_{i0}}{|\bp_j-\bp_i|}$ etc.\
do not vanish. 
Hence, one cannot recover the static limit from the asymptotic
expressions. The expansion does hold, however, in the dynamical limit,
where energy/momentum ratios tend to infinity.
\par\smallskip
Using the power-series expansion
$\ln(1-z) = - \sum_{n=1}^{\infty} \frac{1}{n} z^n$,
we can expand
\begin{equation}\label{eq30}
 \ln\left(\frac{w_{ij}(1)}{w_{ij}(0)}\right) =
 \left.
 \ln\left[ x_{ij}(s) - i \bar y_{ij} \right] -
 \sum_{n=1}^{\infty} \frac{1}{n} 
 \left[ \frac{\bar x_{ij} - i y_{ij}(s)}
             {x_{ij}(s) - i \bar y_{ij}} \right]^n \>
 \right|_{s=0}^{s=1}
\end{equation}
The n-th term in the sum is of order $\lam^n$.
To compute $I_3$, we have to sum over all pairs $(i,j)$ with 
$i,j \in \{1,2,3\}$ and $i \neq j$.
Since
\begin{equation}\label{eq31}
 \bar x_{ji} = - \bar x_{ij} \>, \>
 \bar y_{ji} = - \bar y_{ij} \quad {\rm and} \quad
  x_{ji}(s) = - x_{ij}(s) \>, \>
  y_{ji}(s) = y_{ij}(s) \>,
\end{equation}
the first term in (\ref{eq30}) cancels when the contributions from
$\gam_{ij}$ and $\gam_{ji}$ are subtracted (see Eq.\ (\ref{eq28})), i.e.
\begin{equation}\label{eq32}
 \int_{\gam_{ij}} \frac{dz}{z} - \int_{\gam_{ji}} \frac{dz}{z} =
 \sum_{n=1}^{\infty} \frac{1}{n} \left.
 \left\{ \left[ \frac{\bar x_{ij} + i y_{ij}(s)}
             {x_{ij}(s) - i \bar y_{ij}} \right]^n 
       - \left[ \frac{\bar x_{ij} - i y_{ij}(s)}
             {x_{ij}(s) - i \bar y_{ij}} \right]^n \right\} \>
 \right|_{s=0}^{s=1}
\end{equation}
The leading term for small $\lam$ is the one with $n = 1$.
Neglecting the other terms, one obtains after some elementary algebra
(see Appendix A),
\begin{eqnarray}\label{eq33}
 I_3(p_1,p_2,p_3) \to \lam^{-1} 
 \frac{1}{2\pi i \, \det(\bp_2\!-\!\bp_1,\bp_3\!-\!\bp_1)} \,
 \frac{1}{1 + (\Im\,\bd)^2} \nonumber \\
 \times 
 \sum_{(i,j) = (1,2),(2,3),(3,1)} 
 \frac{\det(\Im\,\bd,\bp_j\!-\!\bp_i)}
      {\sqrt{1 + |\bp_j\!-\!\bp_i|^2/(p_{j0}\!-\!p_{i0})^2}}
\end{eqnarray}
with corrections of order one. Hence, a single 3-loop generally diverges 
as $\lam^{-1}$ for $\lam \to 0$, just as expected from simple 
power-counting.
\begin{figure}
\epsfbox{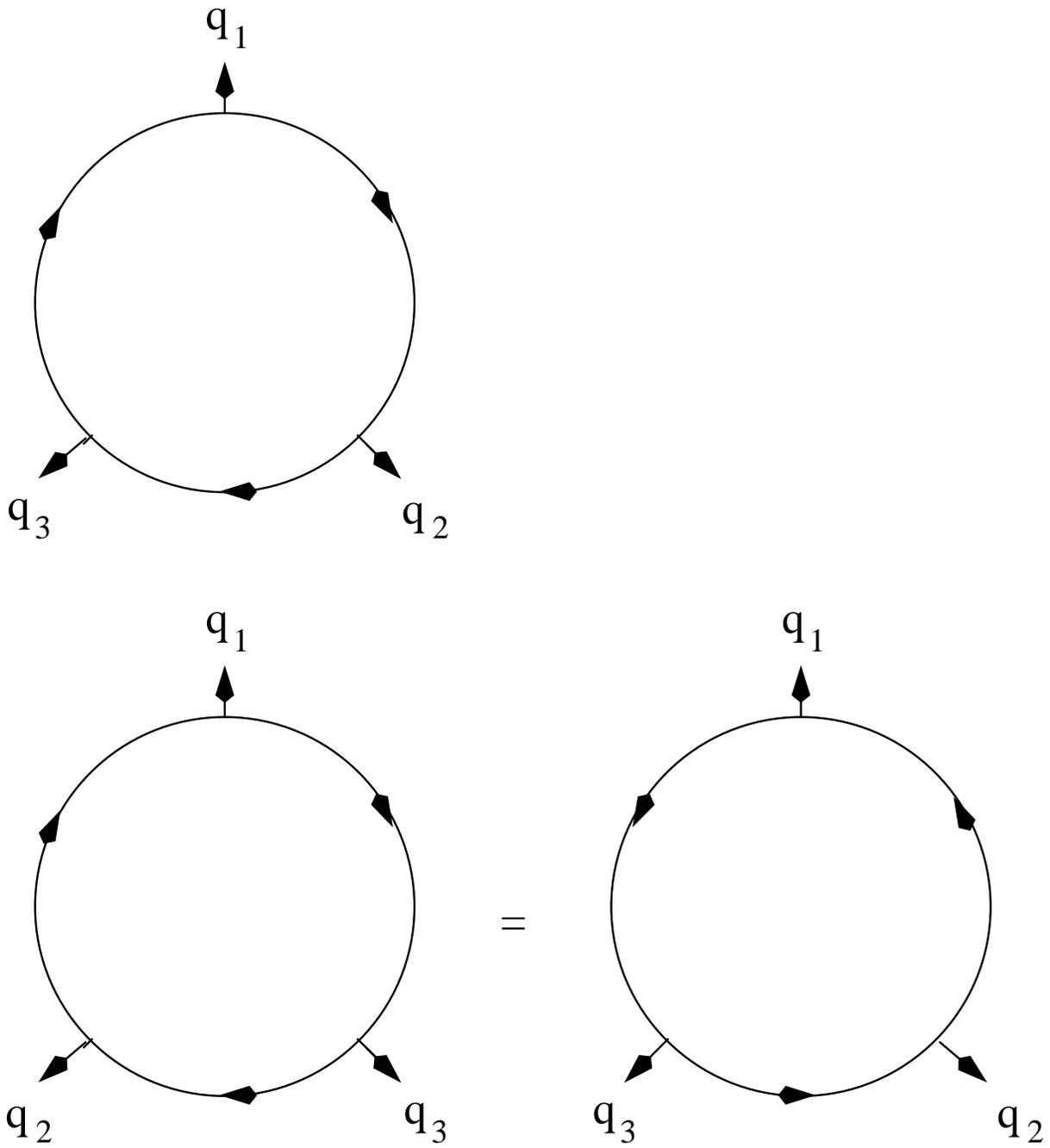}
\caption{The two distinct permutations for the 3-loop.}
\end{figure}
\par\smallskip
We now determine the symmetrized 3-loop for small energies
and momenta.
There are only two non-equivalent permutations of $q_1,q_2,q_3$
(see Fig.\ 2), corresponding to a change of sign of $p_1,p_2,p_3$.
The symmetrized loop $\Pi_3^S$ can therefore be expressed as
\begin{equation}\label{eq34}
 \Pi_3^S(q_1,q_2,q_3) = 
 \frac{1}{2} \left[ I_3(p_1,p_2,p_3) + I_3(-p_1,-p_2,-p_3) \right] 
 \equiv I_3^S(p_1,p_2,p_3)
\end{equation}
It is easy to see that
\begin{equation}\label{eq35}
 \bar x_{ij} \mapsto   \bar x_{ij} \> , \>
 \bar y_{ij} \mapsto - \bar y_{ij} \quad {\rm and} \quad
 x_{ij}(s)   \mapsto - x_{ij}(s) \> , \>
 y_{ij}(s)   \mapsto   y_{ij}(s) \quad
 {\rm for} \quad p_i \mapsto - p_i.
\end{equation}
Hence, the first order terms cancel when summing permuted loops, 
and the leading contributions come from $n=2$ in the expansion 
(\ref{eq32}), i.e.\
\begin{equation}\label{eq36}
 \Pi_3^S(q_1,q_2,q_3) \to
 \frac{1}{4 \pi i \, \det(\bp_2\!-\!\bp_1,\bp_3\!-\!\bp_1)} \,
 \sum_{(i,j) = (1,2),(2,3),(3,1)} \, \left. \left[
 \left( X_{ij}^+(s) \right)^2 - \left( X_{ij}^-(s) \right)^2
 \right] \right|_{s=0}^{s=1} \,
\end{equation}
where
\begin{equation}\label{eq37}
 X_{ij}^{\pm}(s) \equiv
 \frac{\bar x_{ij} \pm iy_{ij}(s)}{x_{ij}(s) - i\bar y_{ij}} \, .
\end{equation}
After some simple algebra (see Appendix A) one obtains
\begin{eqnarray}\label{eq38}
 \Pi_3^S(q_1,q_2,q_3) \to
 \frac{1}{\pi \, \det(\bp_2\!-\!\bp_1,\bp_3\!-\!\bp_1)} \,
 \frac{1}{[s^2 + (\Im\,\bd)^2]^2} \, \sum_{(i,j) = (1,2),(2,3),(3,1)}
 \bar x_{ij} \hskip 2cm \nonumber \\
 \times \left. \left[
  [s^2 + (\Im\,\bd)^2] \,
  \frac{|\bp_j\!-\!\bp_i|/2}
       {\sqrt{1 + s^2 |\bp_j\!-\!\bp_i|^2/(p_{j0}\!-\!p_{i0})^2}}
   - |p_{j0}\!-\!p_{i0}| 
  \sqrt{s^2 + \frac{(p_{j0}-p_{i0})^2}{|\bp_j\!-\!\bp_i|^2}} \,
  \right]
 \right|_{s=0}^{s=1}
\end{eqnarray}
with corrections of order $\lam$. The symmetrized loop is thus 
{\em finite}\/ (of order one) for $\lam \to 0$, i.e. the divergence
present in single loops is cancelled when summing permutations.
Note also that $\Pi_3^S(q_1,q_2,q_3)$ is obviously {\em real}\/ in 
the small-q limit.
\par\smallskip
Next we analyze the behavior of the symmetrized 3-loop for {\em one}\/ 
vanishing external momentum variable, say $\bq_1$, while the other
momenta and all energy variables remain finite.
Noting that $\det(\bp_2\!-\!\bp_1,\bp_3\!-\!\bp_1)$ is of order 
$\bq_1$ in that limit, it is easy to see that $X_{ij}^{\pm}(s) = 
O(|\bq_1|)$ for all $i,j$.
Hence, we can use the expansion (\ref{eq32}) once again. For the
symmetrized loop the first order terms cancel, and the leading
contribution is given by (\ref{eq36}), which is of order $|\bq_1|$.
\par\smallskip
We finally consider the dynamical limit, where all momenta scale to
zero (with a scaling factor $\lam$) at fixed finite energy variables.
In that limit $X_{ij}^{\pm}(s)$ is of order $\lam^2$, so that the 
leading contribution to the symmetrized loop is again given by 
(\ref{eq36}).
Solving for $z_{ij}(s)$ for small $\lam$, one can easily show that the 
difference $X_{ij}^{\pm}(1) - X_{ij}^{\pm}(0)$ is only of order
$\lam^4$. Since $\det(\bp_2\!-\!\bp_1,\bp_3\!-\!\bp_1)$ is obviously
of order $\lam^2$ in the dynamical limit, one obtains the result
\begin{equation}\label{eq39}
 \Pi_3^S((q_{10},\lam\bq_1),(q_{20},\lam\bq_2),(q_{30},\lam\bq_3))
 = O(\lam^4) \quad \mbox{for} \quad \lam \to 0
\end{equation}

\vskip 1cm

\noindent
{\large\bf 5. Permutations of external energy-momentum variables} \par
\medskip

Consider a term in the reduction formula (\ref{eq14}) with fixed values 
$i,j,k$.
Let $\{\tq_0,\tq_1,\dots,$ $\tq_M\}$ $\subset$ $\{q_1,\dots,q_N\}$ be the
subset of those external energy-momentum variables that are attached to 
the loop between the lines labelled by $p_j$ and $p_k$, as shown in
Fig.\ 3 (the case of variables attached between $p_i$ and $p_j$ or $p_k$
and $p_i$ is completely analogous).
Define $\tp_0 = p_j$, $\tp_1 = p_{j+1}$, $\tp_2 = p_{j+2},\dots,$
$\tp_M = p_{k-1}$, $\tp_{M+1} = p_k$.
\par\smallskip
Our aim is to show that the sum
\begin{equation}\label{eq40}
 \tS_M(\tq_0,\tq_1,\dots,\tq_M) \equiv
 \sum_{\tP} \prod_{\mu=1}^M \frac{1}{f^{\tP}_{\mu}}
\end{equation}
over all permutations $\tP$ of $\tq_0,\dots,\tq_M$ is of order one in 
the small-q limit, and vanishes with a power $2M$ in the dynamical limit. 
Here
\begin{equation}\label{eq41}
 f^{\tP}_{\mu} \equiv f^{\tP}_{i,j+\mu}(\bd^{ijk}) =
 (\tbp^{\tP}_{\mu} - \bp_i) \cdot \bd^{ijk} + 
 \frac{1}{2}[\bp_i^2 - (\tbp^{\tP}_{\mu})^2] +
 i (p_{i0} - \tp^{\tP}_{\mu 0})
\end{equation}
is constructed with the energy-momentum variable $\tp^{\tP}_{\mu}$ 
on the $\mu$-th line resulting from the permutation $\tP$.
Note that for a fixed permutation $\tP$, the product 
$\prod_{\mu=1}^M (1/f^{\tP}_{\mu})$ diverges as $\lam^{-M}$ for 
$\lam \to 0$. Only the sum over all permutations turns out to be finite
in that limit, due to systematic cancellations.
\par\smallskip
As a first step, we sum over permutations leading to sequences
\begin{equation}\label{eq42}
\tq_1,\tq_2,\dots,\tq_{\alf},\tq_0,\tq_{\alf+1},\dots,\tq_M,
\end{equation}
where $\tq_1,\dots,\tq_M$ follow the order of their indices, while
$\tq_0$ is placed between $\tq_{\alf}$ and $\tq_{\alf+1}$.
We denote these particular permutations by $(\alf)$, where $\alf \in
\{ 0,1,\dots,M \}$. Obviously
\begin{equation}\label{eq43}
 \tp^{(\alf)}_{\mu} = 
 \left\{ \begin{array}{l@{\quad {\rm for} \quad}l}
 \tp'_{\mu} = \tp_{\mu+1} - \tq_0 & \mu \leq \alf \\
 \tp_{\mu} & \mu > \alf
 \end{array} \right.
\end{equation}
Define
\begin{eqnarray}\label{eq44}
 f^{(\alf)}_{\mu} \equiv f^{(\alf)}_{i,j+\mu}(\bd^{ijk}) &=&
 (\tbp^{(\alf)}_{\mu} - \bp_i) \cdot \bd^{ijk} + 
 \frac{1}{2}[\bp_i^2 - (\tbp^{(\alf)}_{\mu})^2] +
 i (p_{i0} - \tp^{(\alf)}_{\mu 0})  \nonumber \\  & \equiv &
 \left\{ \begin{array}{l@{\quad {\rm for} \quad}l}
 f'_{\mu} & \mu \leq \alf \\
 f_{\mu} & \mu > \alf
 \end{array} \right. 
\end{eqnarray}
i.e.\ $f_{\mu}$ is constructed with $\tp_{\mu}$ and $f'_{\mu}$ with
$\tp'_{\mu} = \tp_{\mu} + \tq_{\mu} - \tq_0$.
For the sum over all permutations $(\alf)$ we have derived the 
following \\[2mm]
{\em Lemma 1}\/:
\begin{equation}\label{eq45}
 \sum_{\alf=0}^M \prod_{\mu=1}^M \frac{1}{f^{(\alf)}_{\mu}} =
 \frac{1}{f_1} \sum_{\alf=1}^M 
  \tbq_0 \cdot (\tbq_1 + \dots + \tbq_{\alf})
 \prod_{\mu=1}^M \frac{1}{f^{(\alf)}_{\mu}}
\end{equation}
{\em Proof}\/: Using $\tp'_{\mu} = \tp_{\mu+1} - \tp_1 + p_j$ and
the identity $f_0 \equiv f_{ij}(\bd^{ijk}) = 0$,
it is easy to derive the relation
\begin{equation}\label{eq46}
 f_1 + f'_{\mu} = f_{\mu+1} + g_{\mu+1} 
\end{equation}
where
\begin{equation}\label{eq47} 
 g_{\alf+1} \equiv
 \tbq_0 \cdot (\tbp_{\alf+1}\!-\!\tbp_1) =
 \tbq_0 \cdot (\tbq_1 + \dots + \tbq_{\alf})
\end{equation}
This yields
\begin{equation}\label{eq48}
 \left[ \frac{1}{f_1} + \frac{1}{f'_{\alf}} \right]
 \prod_{\mu=1 \atop \mu \neq \alf}^M \frac{1}{f_{\mu}^{(\alf)}} =
 \frac {f_{\alf+1} + g_{\alf+1}}{f_1} 
 \prod_{\mu=1}^M \frac{1}{f^{(\alf)}_{\mu}}
\end{equation}
We now prove the identity (\ref{eq45}) by starting from the right hand
side. Relation (\ref{eq48}) yields
\begin{equation}\label{eq49}
 \sum_{\alf=1}^M \frac{g_{\alf+1}}{f_1} \prod_{\mu=1}^M 
 \frac{1}{f_{\mu}^{(\alf)}} = 
 \sum_{\alf=1}^M \prod_{\mu=1}^M \frac{1}{f_{\mu}^{(\alf)}} +
 \frac{1}{f_1} \sum_{\alf=1}^M (f'_{\alf} - f_{\alf+1})
 \prod_{\mu=1}^M \frac{1}{f_{\mu}^{(\alf)}}
\end{equation}
The second term reduces to $\prod_{\mu=1}^M (1/f_{\mu})$, due to a 
cancellation of all terms but the first one in the $\alf$-sum 
(note that $f_{M+1} \equiv f_{ik}(\bd^{ijk}) = 0$). 
This completes the proof of Lemma 1.
For energy-momentum variables attached between $p_i$ and $p_j$ or 
between $p_k$ and $p_i$, it can be proved in the same way.
\begin{figure}
\epsfbox{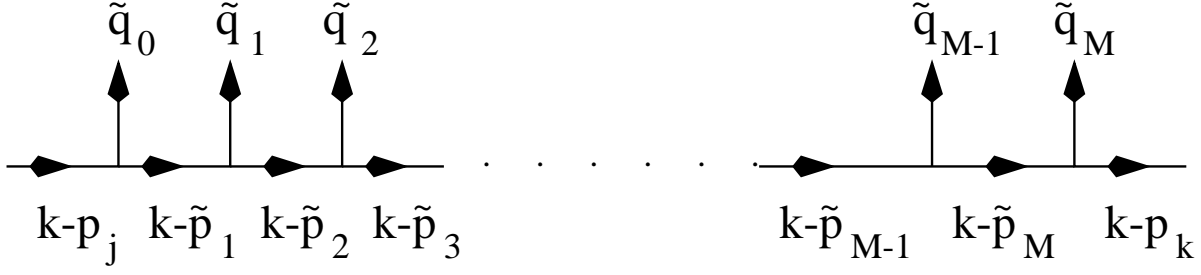}
\caption{Part of an N-loop from \(p_j\) to \(p_k\).}
\end{figure}
\par\smallskip
For the following formal manipulations it will be helpful to represent
products of inverse $f$-factors $\prod_{\mu=1}^M (1/f^{\tP}_{\mu})$ by
the ordered sequence of momentum labels corresponding to the permutation
$\tP$, i.e.
\begin{equation}\label{eq50}
 \prod_{\mu=1}^M \frac{1}{f^{\tP}_{\mu}} \equiv
 (\tP 0, \tP 1, \tP 2,\dots,\tP M)
\end{equation}
With this notation, Lemma 1 can also be written as
\begin{equation}\label{eq51}
 \sum_{\alf=0}^M 
 (1,2,\dots,\alf,0,\alf+1,\dots,M) =
 \frac{1}{f_1} \sum_{\alf,\rho=1 \atop \rho \leq \alf}^M 
  (\tbq_0 \cdot \tbq_{\rho}) \:
 (1,2,\dots,\alf,0,\alf+1,\dots,M)
\end{equation}
In particular, for $M=1$ one obtains
\begin{equation}\label{eq52}
 (0,1) + (1,0) = \frac{\tbq_0 \cdot \tbq_1}{f_1} \, (1,0)
\end{equation}
and for $M=2$,
\begin{equation}\label{eq53}
 (0,1,2) + (1,0,2) + (1,2,0) =
 \frac{1}{f_1} \left\{ 
 (\tbq_0 \cdot \tbq_1) \, (1,0,2) + 
 [\tbq_0 \cdot (\tbq_1 + \tbq_2)] \, (1,2,0) \right\}
\end{equation}
\par
The sum over permutations $(\alf)$ has reduced the degree of divergence
for $\lam \to 0$ by one power.
We now consider permutations $(\alf,\beta)$ with $\beta \leq \alf$ leading
to sequences
\begin{equation}\label{eq54}
 \tq_2,\tq_3,\dots,\tq_{\beta},\tq_1,\tq_{\beta+1},\dots,\tq_{\alf},\tq_0,
 \tq_{\alf+1},\dots,\tq_M
\end{equation}
i.e.\ $\tq_2,\dots,\tq_M$ follow the order of their indices while
$\tq_1$ is placed at an arbitrary position before $\tq_0$ in the
sequence. The corresponding energy-momentum variables on fermion lines
become
\begin{equation}\label{eq55}
 \tp^{(\alf,\beta)}_{\mu} = 
 \left\{ \begin{array}{l@{\quad {\rm for} \quad}l}
 \tp''_{\mu} = \tp_{\mu+2} - \tq_0 - \tq_1  & \mu \leq \beta \\
 \tp'_{\mu}  = \tp_{\mu+1} - \tq_0          & \beta < \mu \leq \alf \\
 \tp_{\mu} & \mu > \alf
 \end{array} \right.
\end{equation}
and the associated product of inverse f-factors is
\begin{equation}\label{eq56}
 \prod_{\mu=1}^M \frac{1}{f_{\mu}^{(\alf,\beta)}} \> =  
 (2,3,\dots,\beta,1,\beta+1,\dots,\alf,0,\alf+1,\dots,M)
\end{equation}
The sum over all permutations $(\alf,\beta)$ with fixed $\alf$ and
$\beta \leq \alf$, can be rewritten using \\[2mm]
{\em Lemma 2}\/:
\begin{equation}\label{eq57}
 \sum_{\beta=0}^{\alf-1} \prod_{\mu=1}^M 
 \frac{1}{f_{\mu}^{(\alf,\beta)}} \> = \>
 \frac{1}{f'_1} \sum_{\beta=1}^{\alf-1} 
 \tbq_1 \cdot (\tbq_2 + \dots + \tbq_{\beta+1})
 \prod_{\mu=1}^M \frac{1}{f_{\mu}^{(\alf,\beta)}} \> + \>
 \frac{1}{f'_1} \prod_{\mu=1}^{\alf-1} \frac{1}{f''_{\mu}}
 \prod_{\mu=\alf+1}^M \frac{1}{f_{\mu}}
\end{equation}
The second term on the right hand side can be written symbolically as
\begin{equation}\label{eq58}
 \frac{1}{f'_1} \prod_{\mu=1}^{\alf-1} \frac{1}{f''_{\mu}}
 \prod_{\mu=\alf+1}^M \frac{1}{f_{\mu}} =
 \frac{1}{f'_1} (2,3,\dots,\alf,[0,1],\alf+1,\dots,M) 
\end{equation}
where $[0,1]$ stands for the sum of two energy-momentum variables, 
$\tq_0 + \tq_1$, i.e.\ the product of inverse f-factors above corresponds
to the sequence $\tq_2,\tq_3,\dots,\tq_{\alf},\tq_0 + \tq_1,\tq_{\alf+1},
\dots,\tq_M$.
\par\smallskip
\noindent
{\em Proof}\/: The proof of Lemma 2 is almost identical to the one for
Lemma 1. 
A difference arises only in the last step, where now the "boundary term"
(\ref{eq58}) contributes, because $f_{\alf} \neq 0$ replaces the factor
$f_{M+1} = 0$ (see Eq.\ (\ref{eq49}) and below).
\par\smallskip
\noindent
For example, for $M=2$, $\alf=1$, Lemma 2 yields the trivial identity
\begin{equation}\label{eq59}
 (1,0,2) = \frac{1}{f'_1} \, ([0,1],2) 
\end{equation}
and for $M=2$, $\alf=2$,
\begin{equation}\label{eq60}
 (1,2,0) + (2,1,0) = 
 \frac{1}{f'_1} \, (\tbq_1 \cdot \tbq_2) \, (2,1,0) +
 \frac{1}{f'_1} \, (2,[0,1])
\end{equation}
An example for $M=3$ can be found in Appendix B.
\par\smallskip
Lemma 1 and Lemma 2 have been formulated and derived for $\tq_0$ and
$\tq_1$ "running" from left to right, with a specific fixed order of 
the other variables. Analogous identities hold of course for any
permutation of running and fixed variables, e.g.\ for $\tq_3$ running
with some fixed order of all other variables.
In general, if $\tq_{\rho}$ is the running variable, the explicit
prefactor $1/f_1$ in Lemma 1 and $1/f'_1$ in Lemma 2 has to be 
replaced by a factor $1/h_{\rho}$, where 
\begin{equation}\label{eq61}
 h_{\rho} = (\bp_j + \tbq_{\rho} - \bp_i) \cdot \bd^{ijk} +
 \frac{1}{2} [\bp_i^2 - (\bp_j + \tbq_{\rho})^2] +
 i [p_{i0} - (p_{j0} + \tq_{\rho 0})]  
\end{equation}
i.e.\ an $f$-function constructed with $p_i$ and $p_j + \tq_{\rho}$.
Note that $f_1 = h_0$ and $f'_1 = h_1$ by definition.
\par\smallskip
We are now ready to prove the main result of this section: \\[2mm]
{\em Theorem 1}\/: 
{\em The sum $\tS_M(\tq_0,\tq_1,\dots,\tq_M) =
\sum_{\tP} \prod_{\mu=1}^M \frac{1}{f_{\mu}^{\tP}}$ can
be written as a sum over fractions with numerators}
\begin{equation}\label{eq62}
 (\tbq_{\sg_1} \cdot \tbq_{\sg'_1}) \; 
 (\tbq_{\sg_2} \cdot \tbq_{\sg'_2}) \; \dots \;
 (\tbq_{\sg_M} \cdot \tbq_{\sg'_M})
\end{equation}
{\em where $\sg_i \neq \sg'_i$,
and denominators}
\begin{equation}\label{eq63}
 h_0 \, h_1 \, \dots \, h_M \, \times \prod_1^{M-1} f \mbox{-factors}  
\end{equation}
{\em where the $f$-factors are constructed with $p_i$ and $p_j$ + some 
variables $\tq$. In each numerator each momentum variable from 
$\{ \tbq_0,\dots,\tbq_M \}$ appears at least once.}
\par\smallskip
\noindent
{\em Proof}\/: The following {\em algorithm}\/ reduces the sum
$\sum_{\tP} \prod_{\mu=1}^M \frac{1}{f_{\mu}^{\tP}}$ 
to the form described in the theorem.
We proceed by induction with respect to the number of energy-momentum
variables $M+1$.
We will prove that the algorithm works for the case $M=1$ directly,
and for general $M$ under the assumption that it works for less than
$M+1$ variables. 
\\[2mm]
{\em Step 1}\/: Write down all permutations of $(0,1,\dots,M)$, and
apply Lemma 1 to groups with a fixed order of $1,\dots,M$, and $0$
running over all possible positions. 
This yields
\begin{equation}\label{eq64}
 \sum_{\tP} \prod_{\mu=1}^M \frac{1}{f_{\mu}^{\tP}} =
 \sum_{\rho=1}^M
 \frac{\tbq_0 \cdot \tbq_{\rho}}{h_0} \: 
 \sum \, (\dots,\rho,\dots,0,\dots) 
\end{equation}
where, for fixed $\rho$, all products 
$(\dots,\rho,\dots,0,\dots)$ with $\rho$ to the left of $0$ contribute 
exactly once.
For $M=1$ one gets simply
\begin{equation}\label{eq65}
 \tS_1(\tq_0,\tq_1) = (0,1) + (1,0) =
 \frac{\tbq_0 \cdot \tbq_1}{h_0} \: (1,0) = 
 \frac{\tbq_0 \cdot \tbq_1}{h_0 \, h_1}  
\end{equation}
i.e.\ for this case Theorem 1 is already proven. 
\par\smallskip
\noindent
{\em Step 2}\/: The coefficients $\sum \, (\dots,\rho,\dots,0,\dots)$
of $\frac{\tbq_0 \cdot \tbq_{\rho}}{h_0}$ can be further reduced by
applying Lemma 2 to groups with fixed positions of all variables
except $\rho$ and $\rho$ running from the first place to the place 
before and nearest to $0$. This yields a sum
\begin{equation}\label{eq66}
 \sum_{\rho' = 1 \atop \rho' \neq 0,\rho}^M 
 \frac{\tbq_{\rho} \cdot \tbq_{\rho'}}{h_{\rho}} \:
 \sum \, (\dots,\rho',\dots,\rho,\dots,0,\dots) +
 \frac{1}{h_{\rho}} \: 
 \sum \, (\dots,[0,\rho],\dots) 
\end{equation}
where the coefficient $\sum \, (\dots,\rho',\dots,\rho,\dots,0,\dots)$
is a sum over all permutations with $\rho'$ to the left of $\rho$ 
and $\rho$ to the left of $0$. 
The "boundary term" $\sum \, (\dots,[0,\rho],\dots)$ is a sum over all
permutations of products of $M\!-\!1$ $f$-factors with $M$ 
energy-momentum variables, i.e.\ all variables
$\tq_{\mu}$ with $\mu \neq 0,\rho$ and $\tq_0 + \tq_{\rho}$.
According to our induction hypothesis we can apply Theorem 1 to this
term, which, with the prefactor 
$\frac{\tbq_0 \cdot \tbq_{\rho}}{h_0 h_{\rho}}$ obtained in Step 1
and Step 2, leads to a sum over fractions with the required structure.
\par\smallskip
\noindent
{\em Step 3}\/: To reduce the first sum of terms obtained in Step 2,
we apply Lemma 2 to the coefficients 
$\sum \, (\dots,\rho',\dots,\rho,\dots,0,\dots)$, this time with
$\rho'$ running from the first place to $\rho$. This yields sums
\begin{equation}\label{eq67}
 \sum_{\rho'' = 1 \atop \rho'' \neq 0,\rho,\rho'}^M 
 \frac{\tbq_{\rho'} \cdot \tbq_{\rho''}}{h_{\rho'}} \:
 \sum \, (\dots,\rho'',\dots,\rho',\dots,\rho,\dots,0,\dots) +
 \frac{1}{h_{\rho'}} \: 
 \sum \, (\dots,[\rho,\rho'],\dots,0,\dots) \quad
\end{equation}
Again, the boundary term leads back to a case with one variable
less, which is reducible to the desired form by virtue of the 
induction hypothesis.
\par\smallskip
\noindent
{\em Steps 4-M}\/: The first sum of terms from Step 3 can be reduced 
by another application of Lemma 2, and so on. The boundary terms
always lead back to a case with a reduced number of energy-momentum
variables. The algorithm terminates after the $M$-th step, where 
coefficients $(\rho^{(M-1)},\rho^{(M-2)},\dots,\rho'',\rho',\rho,0)$ 
with a fixed order appear, where $\rho^{(M-1)}$ is the only variable 
that has not "run" (in the sense of Lemma 2). These coefficients 
correspond to inverse products of $M$ $f$-factors, the first of which
is $h_{\rho^{(M-1)}}$.
It is clear that each momentum variable appears at least once in each
product of $M$ scalar products generated by the algorithm.
This completes the proof of the theorem.
\par\smallskip
For $M=2$, for example, the algorithm yields
\begin{equation}\label{eq68}
 \tS_2(\tq_0,\tq_1,\tq_2) \: = \:
 (\tbq_0 \cdot \tbq_1) \: (\tbq_1 \cdot \tbq_2) \:
 \frac{1}{h_0 \, h_1 \, h_2} \, 
 \left[ \frac{1}{h_{0,1}} + \frac{1}{h_{1,2}} \right] 
 \: + \: \mbox{cyclic permutations}
\end{equation}
where $h_{\rho,\rho'}$ is an $f$-function constructed with $p_i$ and
$p_j + \tq_{\rho} + \tq_{\rho'}$.
In Appendix B the algorithm is presented at work for the case $M = 3$.
\par\smallskip
The following important corollaries follow directly from Theorem 1. 
\\[2mm]
{\em Corollary 1}\/: \\ 
$\tS_M(\tq_0,\tq_1,\dots,\tq_M)$ is {\em finite}\/ (of
order one) and {\em real}\/ in the small-q limit, where all momentum
and energy variables scale to zero as $\lam$.
\\[2mm]
{\em Corollary 2}\/: \\
$\tS_M(\tq_0,\tq_1,\dots,\tq_M)$ is of order 
$\tbq_{\rho}$ for $\tbq_{\rho} \to 0$ if $\tq_{\rho 0}$ and all 
$\tq_{\mu}$ with $\mu \neq \rho$ remain finite.
\\[2mm]
{\em Corollary 3}\/: \\
In the dynamical limit, where all momenta scale to zero as $\lam$ at
fixed finite energy variables, the sum $\tS_M(\tq_0,\tq_1,\dots,\tq_M)$ 
vanishes as $\lam^{2M}$. 
\\[2mm]
{\em Proof}\/: Theorem 1 tells us that $\tS_M(\tq_0,\tq_1,\dots,\tq_M)$
is a sum of fractions with $M$ scalar products of momenta in the 
numerator and $2M$ $f$-factors in the denominator. The $f$-factors
vanish linearly in the small-q limit (where momenta and energies tend
both to zero), and are purely imaginary to leading order in $\lam$.
This yields Corollary 1. 
The $f$-factors remain finite if momenta vanish at finite energies.
This yields the Corollaries 2 and 3, where for the former the observation
that in each numerator each momentum variable appears at least once
is crucial.
\par\smallskip
Actually Corrollary 2 holds already for a sum over all positions of
$\tq_{\rho}$ with fixed order of the other variables (instead of 
summing all permutations), and follows directly from Lemma 1. 
\par\smallskip
In the following sections we will use these results to control the
behavior of symmetrized N-loops in the above limits.
\par

\vskip 1cm

\noindent
{\large\bf 6. The N-loop} \par
\medskip

Any N-loop can be computed by inserting the explicit expression for the
3-loop given in Sec.\ 4 into the reduction formula (\ref{eq14}),
with $f_{i\nu}(\bd^{ijk})$ from (\ref{eq17}). 
\par\smallskip
In the small-q limit, where all momenta and energy variables scale to
zero as $\lam$, the unsymmetrized N-loop diverges as $\lam^{2-\N}$,
since $I_3$ diverges as $\lam^{-1}$ and the product of $\N\!-\!3$ factors
$f_{i\nu}(\bd^{ijk})$ vanishes as $\lam^{\N-3}$.
We will now show that the symmetrized N-loop (\ref{eq7}) remains finite in
the small-q limit.
\par\smallskip
Symmetrizing the reduction formula (\ref{eq14}) for the N-loop, one can
write
\begin{equation}\label{eq69}
 \Pi_\N^S(q_1,\dots,q_\N) \: = \: \cS
 \sum_{1 \leq i < j < k \leq N} S_{ij} \, S_{jk} \, S_{ki} \,
 I_3(p_i,p_j,p_k)
\end{equation}
Here $S_{ij}$, $S_{jk}$ and $S_{ki}$ are symmetrized products of
inverse $f$-factors, e.g. 
\begin{equation}\label{eq70}
 S_{jk} \equiv \left\{ \begin{array}{l@{\quad {\rm for} \quad}l}
 \frac{1}{(M+1)!} \, \tS_M & M = k-j-1 \geq 1 \\
 1                         & k = j+1
 \end{array} \right.
\end{equation}
with $\tS_M$ as defined in Sec.\ 5. Note that symmetrizing once or twice, 
or first symmetrizing partially (with respect to a subset of variables) 
and then completely (by applying $\cS$), or vice versa, always yields the 
same result.
In particular,
\begin{equation}\label{eq71}
 \Pi_\N^S(q_1,\dots,q_\N) = \frac{1}{2} \, \left[
 \Pi_\N^S(q_1,\dots,q_\N) + (p_1,\dots,p_\N) \mapsto (-p_1,\dots,-p_\N)
 \right] 
\end{equation}
Since $f_{i\nu}(\bd^{ijk}) \mapsto f^*_{i\nu}(\bd^{ijk})$ and thus
$\tS_M \mapsto \tS_M^*$ for 
$(p_1,\dots,p_\N) \mapsto (-p_1,\dots,-p_\N)$, 
one can also write
\begin{equation}\label{eq72}
 \Pi_\N^S(q_1,\dots,q_\N) \: = \: \cS
 \sum_{1 \leq i < j < k \leq N} 
 \left[ \,
      \Re (S_{ij} S_{jk} S_{ki}) \, I_3^S(p_i,p_j,p_k) +
 i \, \Im (S_{ij} S_{jk} S_{ki}) \, I_3^A(p_i,p_j,p_k)
 \right] 
\end{equation}
where $I_3^A(p_i,p_j,p_k) \equiv 
\frac{1}{2} \, I_3(p_i,p_j,p_k) - I_3(-p_i,-p_j,-p_k)$.
Note that unsymmetrized loops are complex for $\N > 2$, while
symmetrized loops are {\em real}, since 
$I_\N(-p_1,\dots,-p_\N) = I_\N^*(p_1,\dots,p_\N)$.
\par\smallskip
In the small-q limit, $\Re (S_{ij} \, S_{jk} \, S_{ki})$ is of order
one, while $\Im (S_{ij} \, S_{jk} \, S_{ki})$ is of order $\lam$ (see
Theorem 1 and Corollary 1 in Sec.\ 5).
In Sec.\ 4 we have shown that $I_3^S$ is of order one and real in 
that limit, while $I_3^A$ diverges as $\lam^{-1}$ with a purely
imaginary coefficient (see Eq.\ (\ref{eq33})).
Hence $\Pi_\N^S(q_1,\dots,q_\N)$ is {\em finite}\/ in the small-q 
limit, i.e.
\begin{equation}\label{eq73}
 \Pi_\N^S(\lam q_1,\dots,\lam q_\N) = O(1) \quad
 \mbox{for} \quad \lam \to 0
\end{equation}
\par\smallskip
We now analyse the dynamical limit, where all momenta scale to
zero as $\lam$ at finite energy variables. 
In that limit $\Re f_{i\nu}(\bd^{ijk})$ is of order $\lam^2$, while
$\Im f_{i\nu}(\bd^{ijk})$ is of order one (see, for example, Eq.\
(\ref{eq17}). 
Theorem 1 then implies that $\Re (S_{ij} S_{jk} S_{ki})$ vanishes as 
$\lam^{2(\N-3)}$, and $\Im (S_{ij} S_{jk} S_{ki})$ as $\lam^{2(\N-2)}$.
The symmetrized 3-loop has been shown to vanish as $\lam^4$ in the 
dynamical limit in Sec.\ 4.
Since $X_{ij}^{\pm}(s)$ is of order $\lam^2$ and
$X_{ij}^{\pm}(1) - X_{ij}^{\pm}(0)$ of order $\lam^4$ (see Sec.\ 4),
it is obvious from (\ref{eq32}) that $I_3^A$ vanishes as $\lam^2$ in 
the dynamical limit.
Inserting these asymptotic power-laws in (\ref{eq72}), one finds
\begin{equation}\label{eq74}
 \Pi_\N^S((q_{10},\lam\bq_1),\dots,(q_{\N 0},\lam\bq_\N) = 
 O(\lam^{2\N-2}) \quad \mbox{for} \quad \lam \to 0
\end{equation}
\par\smallskip
We finally discuss the behavior of loops for a single vanishing
momentum, say $\bq_1$, while energy variables and other momenta
remain finite.
Note that this limit cannot be taken for 2-loops, since momentum
conservation imposes $\bq_2 = - \bq_1$ in that case, i.e.\ $\bq_2$
cannot remain finite if $\bq_1$ tends to zero.
Unsymmetrized loops with $\N \geq 3$ remain finite in the small 
$\bq_1$-limit, since $f$-factors and 3-loops remain finite if the
other variables remain finite.
We have already seen in Sec.\ 4 that the symmetrized 3-loop vanishes
linearly for $\bq_1 \to 0$. 
This behavior holds also for the symmetrized N-loop, i.e.\
\begin{equation}\label{eq75}
 \Pi_{\N}^S(q_1,\dots,q_{\N}) = O(|\bq_1|) 
 \quad \mbox{for} \quad \bq_1 \to 0
\end{equation}
This latter result can be established quite easily in any dimension 
by using the techniques reviewed in Refs.\ \cite{MCD,Kop},
for example via Ward identities \cite{CDM,Tru}.
\par\smallskip
The above results for the symmetrized loops have been checked 
numerically for $\N = 3,4,\dots,7$. 
No cancellations beyond the order established here have been observed.
\par

\vskip 1cm

\noindent
{\large\bf 7. Conclusion} \par
\medskip

In summary, we have derived explicit formulae for fermion loops in two
dimensions from an exact expression obtained recently by Feldman et al.\
\cite{FKST}.
The 3-loop is an elementary function of momenta and frequencies, and
the N-loop can be expressed as a linear combination of 3-loops with 
coefficients that are rational functions of momentum and frequency
variables. 
These formulae are very useful for the evaluation of Feynman diagrams
for two-dimensional interacting Fermi systems where such loops appear 
as subdiagrams.
\par\smallskip
We have also analyzed to what extent divergencies in the low energy
and small momentum limit cancel in the symmetrized loop, defined as a
sum over permutations of momentum and frequency variables.
A single loop was seen to diverge with a power $\N-2$ as expected 
from a simple scaling analysis.
We have proved a reduction of the degree of divergence in the 
symmetrized N-loop by $\N-2$ powers, such that symmetrized loops and
thus all N-point density correlation functions of the two-dimensional 
Fermi gas are generally finite in the low energy and small momentum 
limit.
This drastic cancellation justifies the frequent neglect of loops with 
more than two insertions assumed in Ward identity or bosonization methods
\cite{MCD,Kop} for systems where interactions with small momentum
transfers dominate the low-energy physics.  
In the dynamical limit, where momenta scale to zero at finite energy
variables, we have shown that the symmetrized N-loop vanishes as the
$(2\N\!-\!2)$-th power of the scale parameter.
\par\smallskip

\vskip 1cm


\noindent{\bf Acknowledgements:} \\
We are grateful to C. Halboth, H. Kn\"orrer and E. Trubo\-witz for 
valuable discussions. 

\vskip 2cm


\noindent
{\bf Appendix A: 3-loop for small energies and momenta} \par
\medskip

In this appendix we present the algebra leading to the asymptotic
results for the 3-loop in Sec.\ 4.
\par\smallskip
The first order term in (\ref{eq32}) is
\begin{equation}\label{A1}
 \left. \frac{2i \, y_{ij}(s)}{x_{ij}(s) - i \bar y_{ij}} 
   \right|_{s=0}^{s=1} = 
   2i \left. 
   \frac{[x_{ij}(s) + i \bar y_{ij}] \, y_{ij}(s)}
        { x_{ij}^2(s) + \bar y_{ij}^2} \right|_{s=0}^{s=1}
\end{equation}

The denominator can be simplified to
\begin{equation}\label{A2}
 x_{ij}^2(s) + \bar y_{ij}^2 =
 x_{ij}^2(s) + (\Im\,\bd)^2 -
 \left( \Im\,\bd \cdot \frac{\bp_j\!-\!\bp_i}{|\bp_j\!-\!\bp_i|}
 \right)^2 = 
 s^2 + (\Im\,\bd)^2
\end{equation}
where in the second step we have used the fact that $\bd$ is a
zero of the function $f_{ij}(\bk)$ defined in (\ref{eq10}).
Relation (\ref{eq24}) yields the identity
\begin{equation}\label{A3}
 \sum_{(i,j) = (1,2),(2,3),(3,1)} x_{ij}(s) \, y_{ij}(s) = 0
\end{equation}
and 
$\bar y_{ij} \, y_{ij}(0) = \frac{1}{2} \det(\Im\,\bd,\bp_j\!-\!\bp_i)$
yields
\begin{equation}\label{A4}
 \sum_{(i,j) = (1,2),(2,3),(3,1)} \bar y_{ij} \, y_{ij}(0) = 0
\end{equation}
Inserting the above relations in the first order term in (\ref{eq32}) 
one immediately obtains the result (\ref{eq33}).
\par\smallskip
The second order term in (\ref{eq32}) is
\begin{equation}\label{A5}
 \left. 
 \frac{2i \, \bar x_{ij} \, y_{ij}(s)}{[x_{ij}(s) - i \bar y_{ij}]^2}
 \right|_{s=0}^{s=1} = 
 \left. 2i \, 
 \frac{\bar x_{ij} \, y_{ij}(s) 
       [x^2_{ij}(s) - \bar y_{ij}^2 + 2i x_{ij}(s) \, \bar y_{ij}]}
      {[s^2 + (\Im\,\bd)^2]^2} \right|_{s=0}^{s=1}
\end{equation}
where the last step follows from (\ref{A2}). 
Using Eq.\ (\ref{eq24}), i.e. 
$x_{ij}(s) \, y_{ij}(s) = - \frac{1}{2}(p_{j0} - p_{i0})$,
and the identity
\begin{equation}\label{A6}
 \sum_{(i,j) = (1,2),(2,3),(3,1)} \bar x_{ij} \, \bar y_{ij} \>
  (p_{j0} - p_{i0}) = 0
\end{equation}
one obtains the result (\ref{eq38}).
\par

\vskip 1cm

\noindent
{\bf Appendix B: Permutations for $M=3$} \par
\medskip

In this Appendix we carry out the permutation algorithm constructed in
Sec.\ 5 for the case $M=3$.
\par\smallskip
\noindent
{\em Step 1}\/: Write down all permutations of $(0,1,2,3)$, i.e.\ 24
terms, and apply Lemma 1 to groups with fixed order of $1,2,3$.
This yields 
\begin{equation}\label{B1}
 \frac{\tbq_0 \cdot \tbq_1}{h_0} \: \Big[
 (1,0,2,3) + 
 (1,2,0,3) + (2,1,0,3) + 
 (1,2,3,0) + (2,1,3,0) + (2,3,1,0) \: + \:
 2 \lra 3 \Big] \quad
\end{equation}
and analogous terms proportional to $\frac{\tbq_0 \cdot \tbq_2}{h_0}$
and $\frac{\tbq_0 \cdot \tbq_3}{h_0}$.
\par\smallskip
\noindent
{\em Step 2}\/: Applying Lemma 2, the above coefficients of 
$\frac{\tbq_0 \cdot \tbq_1}{h_0}$ can be rewritten as
\begin{eqnarray}\label{B2}
 (1,0,2,3) &=& 
 \frac{1}{h_1} \, ([0,1],2,3)  \nonumber \\
 (1,2,0,3) + (2,1,0,3) &=&
 \frac{\tbq_1 \cdot \tbq_2}{h_1} \, (2,1,0,3) + 
 \frac{1}{h_1} \, (2,[0,1],3)  \nonumber \\
 (1,2,3,0) + (2,1,3,0) + (2,3,1,0) &=&
 \frac{\tbq_1 \cdot \tbq_2}{h_1} \, (2,1,3,0) + 
 \frac{\tbq_1 \cdot (\tbq_2 + \tbq_3)}{h_1} \, (2,3,1,0)  \nonumber \\
 & & + \frac{1}{h_1} \, (2,3,[0,1])
\end{eqnarray}
and analogously for permutations $2 \lra 3$. Applying Lemma 1 to the 
"boundary terms" yields
\begin{eqnarray}\label{B3}
 \frac{1}{h_1} \: \Big[ ([0,1],2,3) + (2,[0,1],3) + (2,3,[0,1]) \Big] 
 &=& \nonumber \\
 \frac{(\tbq_0 + \tbq_1) \cdot \tbq_2}{h_1 h_{0,1}} \: 
 (2,[0,1],3) 
 &+&
 \frac{(\tbq_0 + \tbq_1) \cdot (\tbq_2 + \tbq_3)}{h_1 h_{0,1}} \: 
 (2,3,[0,1]) \quad \quad
\end{eqnarray}
and analogously for permutations $2 \lra 3$.
Hence, the coefficient of $\frac{\tbq_0 \cdot \tbq_1}{h_0}$ becomes
\begin{eqnarray}\label{B4}
 \frac{\tbq_1 \cdot \tbq_2}{h_1} \: 
 \Big[ (2,1,0,3) + (2,1,3,0) + (2,3,1,0) + (3,2,1,0) \Big] & &
 \nonumber \\
 + \quad \frac{(\tbq_0 + \tbq_1) \cdot \tbq_2}{h_1 h_{0,1}} \:
 \Big[ (2,[0,1],3) + (2,3,[0,1]) + (3,2,[0,1]) \Big] & + & 2 \lra 3
\end{eqnarray}
The coefficients of $\frac{\tbq_0 \cdot \tbq_2}{h_0}$ and 
$\frac{\tbq_0 \cdot \tbq_3}{h_0}$ can be rewritten similarly by 
applying Lemma 2 to groups with $2$ and $3$ running from the left end
to the position of $0$ with the order of all other variables fixed. 
\par\smallskip
\noindent
{\em Step 3}\/: We finally show, as an example, how another scalar
product of momenta can be extracted from the coefficient
$(2,1,0,3) + (2,1,3,0) + (2,3,1,0) + (3,2,1,0)$ of 
$\frac{\tbq_0 \cdot \tbq_1}{h_0} \, \frac{\tbq_1 \cdot \tbq_2}{h_1}$, 
obtained in the second step above.
Lemma 2 yields
\begin{eqnarray}\label{B5}
 (2,1,0,3) &=& \frac{1}{h_2} \: ([1,2],0,3)  \nonumber \\
 (2,1,3,0) &=& \frac{1}{h_2} \: ([1,2],3,0)  \nonumber \\
 (2,3,1,0) + (3,2,1,0) &=& 
 \frac{\tbq_2 \cdot \tbq_3}{h_2} \: (3,2,1,0) + 
 \frac{1}{h_2} \: (3,[1,2],0)
\end{eqnarray}
Applying Lemma 2 to the boundary terms in (\ref{B5}) yields
\begin{eqnarray}\label{B6}
 \frac{1}{h_2} \: ([1,2],0,3) &=& 
 \frac{1}{h_2 h_{1,2}} \: ([0,1,2],3)  \nonumber \\
 \frac{1}{h_2} \: \Big[ ([1,2],3,0) + (3,[1,2],0) \Big] &=&
 \frac{(\tbq_1 + \tbq_2) \cdot \tbq_3}{h_2 h_{1,2}} \: (3,[1,2],0) +
 \frac{1}{h_2 h_{1,2}} \: (3,[0,1,2]) \quad \quad
\end{eqnarray}
For the boundary terms above, Lemma 1 yields finally
\begin{equation}\label{B7}
 \frac{1}{h_2 h_{1,2}} \: \Big[ ([0,1,2],3) + (3,[0,1,2]) \Big] =
 \frac{(\tbq_0 + \tbq_1 + \tbq_2) \cdot 
 \tbq_3}{h_2 h_{1,2} h_{0,1,2} h_3}
\end{equation}
Thus only terms which contain products of three scalar products of
momenta are left over.
\par

\vfill\eject


\end{document}